\documentclass[conference]{IEEEtran}
\IEEEoverridecommandlockouts
\usepackage{cite}
\usepackage{amsmath,amssymb,amsfonts}
\usepackage{algorithmic}
\usepackage{graphicx}
\usepackage{textcomp}
\usepackage{xcolor}
\usepackage{enumerate}% For roman letters
\usepackage{colortbl}
\usepackage{multirow}
\usepackage[ruled,linesnumbered]{algorithm2e}
\usepackage{wrapfig}
\usepackage{comment}

\begin{document}

\title{ Multidimensional Human Activity Recognition With Large Language Model: A Conceptual Framework
}

\author{\IEEEauthorblockN{Syed Mhamudul Hasan}
\IEEEauthorblockA{\textit{School of Computing, Southern Illinois University, Carbondale, IL, USA}\\
syedmhamudul.hasan@siu.edu} \vspace{-1.02cm}
}

\maketitle

\begin{abstract}

In high-stake environments like emergency response or elder care, the integration of large language model (LLM), revolutionize risk assessment, resource allocation, and emergency responses in Human Activity Recognition (HAR) systems by leveraging data from various wearable sensors. We propose a conceptual framework that utilizes various wearable devices, each considered as a single dimension, to support a multidimensional learning approach within HAR systems. By integrating and processing data from these diverse sources, LLMs can process and translate complex sensor inputs into actionable insights. This integration mitigates the inherent uncertainties and complexities associated with them, and thus enhancing the responsiveness and effectiveness of emergency services. This paper sets the stage for exploring the transformative potential of LLMs within HAR systems in empowering emergency workers to navigate the unpredictable and risky environments they encounter in their critical roles.

\end{abstract}

\section{Introduction}\label{sec:intro}

In the demanding and often perilous world of emergency response, the integration of advanced technologies can significantly enhance operational efficiency and safety. Emergency personnel, including firefighters, law enforcement officers, mountain rescue teams, and nuclear response teams, operate under extreme conditions where the swift availability of accurate information is crucial. Machine learning (ML), and particularly Human Activity Recognition (HAR), stands at the forefront of this technological revolution, offering new ways to support these critical roles. HAR systems utilize a combination of wearable and ambient sensors to detect and interpret human activities like walking, sitting, and running~\cite{WANG2019167}. The data gathered from these sensors is pivotal in understanding and predicting human behavior in real-time, which is essential for emergency scenarios. However, the effectiveness of these systems depends on the quality and integrity of the data collected. Challenges such as data bias and the risk of data poisoning are significant, as they can lead to inaccuracies that might compromise decision-making during critical operations~\cite{Label_Flipping_Data_Poisoning_Attack}.

The multidimensional analysis of ML models, which evaluates their performance not only in terms of predictive accuracy but also across robustness, scalability, and fairness. Such analyses ensure that ML tools are capable of performing under the diverse and unpredictable conditions that emergency workers face. The use of advanced machine learning techniques like Large Language Model (LLM) can further enhance situational awareness and decision-making processes for these workers, enabling them to make informed decisions swiftly and effectively. In the context of HAR, LLM can significantly enhance the processing and understanding of sensor data, translating diverse sensor inputs into actionable insights. By integrating LLM with HAR systems, the analysis can be enriched through advanced pattern recognition and predictive analytics, which are crucial for interpreting complex human activities. This integration enables emergency responders to better assess risks and make informed decisions by providing a deeper, contextual understanding of the environment and activities occurring within it. For example, an LLM could analyze data from wearable sensors to predict potential health risks or required interventions in real-time, thus improving the responsiveness and effectiveness of emergency services. The use of LLM in conjunction with HAR technologies thus holds the potential to transform how emergency scenarios are managed through the augmentation of human decision-making with powerful, AI-driven insights.

\textit{The goal of this paper is to explore the potential and challenges of the unique integration of LLM with HAR systems, aiming to enhance the capabilities of emergency responders in high-stakes environments. We propose a conceptual framework, consisting of four fundamental phases, to develop such integrated approach. Our contribution also includes a systematic discussion of the challenges that must be addressed in the realization of the system.}

\section{Related Work}\label{sec:related}

The primary purpose of HAR is to develop a system capable of recognizing and understanding human activities from data collected by various sensors, such as accelerometers, gyroscopes, or wearable devices~\cite{gupta2022human}. The applications of HAR span across several domains, including healthcare, fitness tracking, sports analytics, ambient assisted living, human-computer interaction, security surveillance, and many more. HAR systems can provide valuable insights into human behavior, monitor physical activity levels, detect abnormal behaviors, assist in rehabilitation programs, and enhance the user experience in interactive systems. Deep neural networks (DNNs) have emerged as powerful models for addressing a wide range of  tasks that includes image classification, speech recognition, natural language processing, and HAR. These networks are characterized by their ability to learn hierarchical representations of data through multiple layers of interconnected nodes, which enables them to capture complex patterns and relationships within the input data and widely used as HAR classification tasks~\cite{9389739}. A LLM is a special type of DNN model designed to understand and generate human-like text across a wide range of topics and domains. These models are trained on large corpus of text data, typically sourced from the internet or other textual sources~\cite{A_Survey_on_Multimodal_Large_Language_Models}. They are widely used for their ability to generate coherent and contextually relevant text given a prompt or input that can be tailored to any domain. Also, it has many adaptive learning features, like fine-tuning and in-context learning (ICL). Fine-tuning of LLM refers to the process of adapting a pre-trained language model to a specific downstream task or domain by further training it on task-specific data. LLM, such as Generative Pre-trained Transformer (GPT) model, are initially trained on large-scale, diverse text corpora using unsupervised learning techniques, which allows these models to be customized for specific tasks, such as activity recognition~\cite{li2023fine}. In-Context Learning (ICL) represents a pivotal advancement in the capabilities of LLM, enhancing their ability to adapt and respond to a wide range of tasks without the need for extensive training or fine-tuning~\cite{rubin2021learning}. This learning paradigm leverages the model's extensive pre-trained knowledge to understand and execute new tasks based on the context provided through prompts~\cite{dong2022survey, liu2021makes}. GPT can learn and solve new tasks by providing ``training'' examples via prompt. Here, the model can perform tasks by conditioning on input and output instances without optimizing any parameters~\cite{xie2021explanation}. These approaches contrast with traditional machine learning methods that train models from scratch on task-specific datasets which allows for more efficient and flexible adaptation without training the whole model on a separate dataset for each task. Similarly, prompt engineering of LLM is the process of designing and crafting task-specific prompts or inputs to guide the generation or behavior of the model for a specific task. It is particularly relevant in scenarios where practitioners can steer the model's behavior toward producing output that aligns with the task's objectives and desired outcomes by providing tailored prompts or inputs.
\vspace{-12pt}
\begin{figure}[t]
    \centering
    \includegraphics[width = 0.47\textwidth]{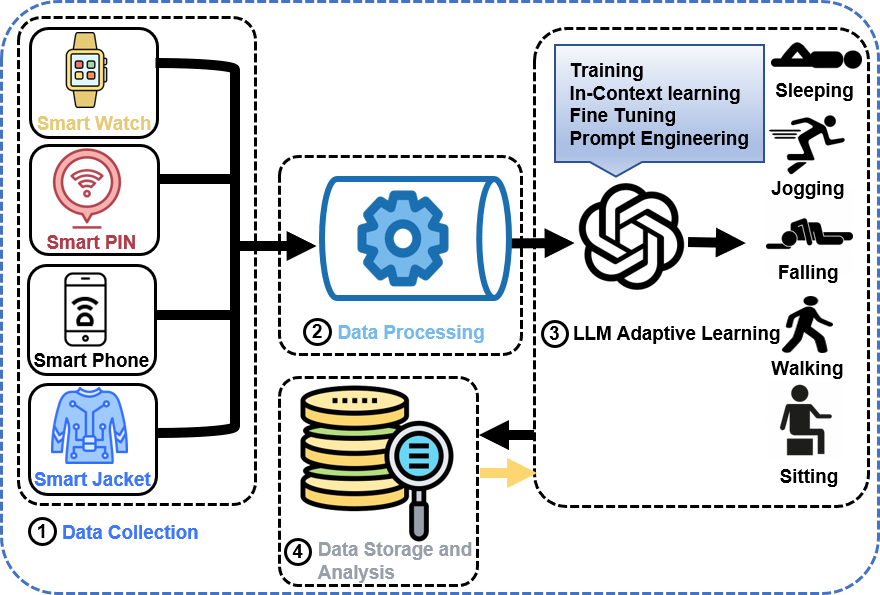}
\vspace{-5pt}
    \caption{\scriptsize The data from the network of sensors embedded in different devices is aggregated through a processing pipeline, enabling a LLM to perform activity recognition. Here, each wearable device is considered a single dimension. 
    }
    \label{fig:har-layers}
\vspace{-18pt}
\end{figure}
\section{System Architecture}

Our proposed multidimensional analysis approach comprises four interdependent submodules. Together, these submodules form a workflow that starts with a raw data collection phase and moves through processing and learning phases to end with storage and graphical visualization. This cohesive workflow ensures the system is both dynamic and responsive to the variety of context-aware activity recognition needs of users and the environments in which it operates.

%\begin{enumerate}[(i)]
\textbf{Data Collection:} The data collection submodule is the foundation of the entire our LLM-integrated HAR system. It allows strategically gathering data from wearable devices positioned across various parts of the body to capture a comprehensive set of human activities and physiological signals~\cite{wan2020deep}. Each type of device and its placement are carefully chosen based on the specific data it can provide. For instance, smartwatches are typically worn on the wrist and particularly effective for monitoring heart rate, activity levels, and sleep patterns. Their placement allows for continuous tracking of cardiovascular health metrics, which are crucial for both fitness monitoring and medical diagnostics. Smartphones, carried in pockets or handbags, are equipped with accelerometers, gyroscopes, and GPS sensors. These devices are invaluable for detecting movement patterns and location-based activities, providing contextual data that enhances the understanding of the user's environment and mobility. On the other hand, wearable smart pins, attached to the clothing at the chest level, are excellent for capturing respiratory rates and detecting changes in posture. Their proximity to the lungs and heart also makes them suitable for monitoring breathing patterns and subtle body movements that may not be captured by wrist-worn devices. Smart jackets incorporate sensors along the arms and torso to monitor body temperature and track upper body movements. These jackets are particularly useful in cold environments and for activities that involve the upper body, such as lifting or throwing, providing data that is critical for ergonomic studies and athletic training. Smart footwear, like sensor-equipped shoes, can collect data on gait, foot pressure distribution, and overall balance. This placement is especially important for applications in sports analytics, rehabilitation, and elderly care, where understanding foot mechanics is crucial for assessing physical health and preventing injuries. By integrating data from devices located across different body parts, our system can achieve a multidimensional understanding of human activities and physiological responses. This holistic view is essential for accurately recognizing complex activities and conditions, improving the overall effectiveness of HAR applications.

%The data collection submodule is the foundation of the entire system. It involves gathering data from various wearable devices such as cellphones, smartwatches, smart jackets, and wearable smart pins. Each device captures specific types of data, which collectively provide a rich, multidimensional dataset. For example, sensor data from a smartphone may include motion and environmental metrics, while a smart watch might collect health-related data. The efficient collection of this diverse data is critical as it feeds into the subsequent processing phase.

%The data collection phase consists of four different mediums: sensor data from the cellphone, smart watch, smart jacket, and wearable smart pin. For cellphones, we have an Android phone with an installed sensor named Sensor Data, which collects data. We gather data from various sources such as the Samsung smart watch, Hexoskin's smart wearable jacket, and Humane's AI-powered wearable smart pin.

\textbf{Data Processing:} The collected data is synchronized and processed in the data processing submodule to ensure that the temporal and spatial disparities between different sensor outputs are aligned. It shapes the raw data collected into a structured form suitable for analysis and decision-making. This involves a sophisticated data pipeline architecture that incorporates both batch and stream processing techniques, optimized for handling the diverse data types collected from various wearable devices. Traditional batch processing involves collecting data in large blocks, which are processed and systematically cleaned, normalized, and transformed to ensure consistency and accuracy before it moves to the next phase of processing. On the other hand, stream processing is essential for real-time applications. It deals with data as it arrives while minimizing latency in information flow. This is crucial for HAR applications requiring immediate feedback, such as emergency alerts for falls in elderly patients or real-time performance adjustments during athletic training. Technologies like Apache Kafka or Apache Flink for real-time data stream processing are employed to manage these flows to ensure dynamic decision-making based on the latest data inputs. As data flows through the pipeline, it is integrated from multiple sources to create a unified view. This process involves aligning data from different sensors and time zones, interpolating missing data points, and enriching the dataset with additional parameters derived from raw data. This enriched dataset is then more effective for training machine learning models as it provides a holistic view of the user’s activities and environmental interactions.

%Data processing is a data pipeline architecture that controls the collection, flow, and delivery of data to storage. This process uses batch processing or stream processing to move data. Batch processing is the legacy approach to the one-time or regularly scheduled movement of data batches from sources to targets without providing for real-time analysis and insights. On the other hand, stream processing enables the real-time movement of data. Stream processing traditionally collects data and change streams from a source such as a database or events from other systems and sensors, which enables real-time business intelligence and decision-making. Thus, we propose to use ETL (Extract, Transform, Load) stream data processing with popular Python ETL libraries like Airflow, PySpark, Luigi, Dask, etc.
%\vspace{-2pt}
\textbf{LLM Adaptive Learning:}  The LLM adaptive learning submodule utilizes the sophisticated capabilities of LLM to meet the specific needs of the HAR system through adaptive learning. This involves a multi-layered approach, starting with model training, where LLM is initially trained using a broad corpus of health data that includes healthcare logs, fitness records, and environmental interactions. These varied data help build a strong foundational understanding of human activities across different contexts. Next, customized prompts can be utilized to guide the LLM in producing accurate and contextually appropriate responses \cite{marvin2023prompt}. For instance, prompts can differentiate activities like running on a flat surface from climbing stairs, enhancing the model's ability to recognize and categorize complex motions from subtle sensor data variations. Fine-tuning is used to further refine the LLM’s capabilities. This approach targets specific HAR tasks with smaller, task-focused datasets to align the model's parameters with the unique characteristics of activities in environments like rehabilitation centers or sports facilities. Equally important, in-context learning (ICL) empowers the model to utilize its broad, pre-trained knowledge base to adapt to new tasks introduced through direct prompts \cite{wei2023larger}. The approach is critical for dynamic environments where activities and conditions change frequently. Additionally, the submodule integrates data from contextual sensors—such as location, time, and weather—with user activity data to deepen the LLM's contextual understanding, thereby enhancing its predictive accuracy regarding the user’s state and intentions. Lastly, the continuous learning component ensures the LLM remains effective over time, constantly learning from new data to refine its understanding and keep pace with evolving user behaviors and emerging activity trends.

\vspace{-2pt}
\textbf{Data Storage and Analysis:} The data storage and analysis submodule ensures that data is not only securely stored but also effectively analyzed to generate actionable insights. This module is equipped with advanced visualization tools and a real-time alerting system to enhance the system's analytical capabilities. The pattern analysis and trend capabilities of the visualization tools allow users to quickly understand and react to the information presented, thereby supporting faster and more informed decision-making. The real-time alerting system plays a pivotal role in operational responsiveness by monitoring specific data parameters and issuing alerts if anomalies or critical conditions are detected. This feature is essential for scenarios where timely intervention is necessary, such as in healthcare monitoring or emergency response situations. Additionally, the historical data and system logs can help understand the nuance and make decision-making more accurate in the future by removing false positives.

\section{Challenges}
The realization of an LLM-integrated HAR system architecture presents several challenges including dataset challenges which denotes collecting and analyzing data from various sensors embedded in different wearable devices (e.g., smart watch, smart pin, smartphone, smart jacket, and so on) presents a unique set of challenges. Each device offers distinct types of data, ranging from health and fitness tracking on a smart watch to motion tracking on a smart pin, which not only captures environmental conditions but also provides valuable motion data through its sensors. Similarly, smartphones contribute a variety of sensor data, like gyroscopic and ambient measurements, while wearable jackets provide detailed cardiac and respiratory metrics. The heterogeneity of data types, combined with the different formats and structures of each source, significantly complicates data integration and processing. Thus, implementing an effective data processing module becomes essential to handling this complex situation. Furthermore, managing large volumes of data necessitates a robust data pipeline, making it critical to use proper ETL (Extract, Transform, Load) libraries. This processing step is critical before training any LLM. Furthermore, ensuring data persistence through reliable storage solutions like SQL or NoSQL DB is crucial for ongoing and future analytical tasks. Each step in this complex data handling process poses its own set of challenges, underscoring the need for advanced techniques and careful planning to achieve accurate and meaningful analysis.

Security and Privacy Issues arises when integrating datasets from diverse sources to extract valuable insights and enhance decision-making processes, the aggregation of data from different sources poses significant security and privacy challenges. Unauthorized access, data breaches, privacy violations, etc. are some of the primary concerns that necessitate security measures and privacy-preserving techniques like federated learning, differential privacy, encryption, etc. Federated learning, a decentralized ML model training architecture, trains the ML model using local data sources without requiring data sharing with a server. As no data sharing is made, data is protected from authorized access~\cite{MOTHUKURI2021619}. Differential privacy provides privacy by adding external noise to the data. By adding controlled noise to data or analysis outputs to protect sensitive attributes from exploitation, thus aligning AI systems by preserving individual data privacy~\cite{8854247}. Encryption can add additional security to data and protect it from security threats. This includes encrypting sensitive data during transmission and storage, implementing access controls and authentication mechanisms to restrict unauthorized access, and protecting the dataset from suspicious activities.

The training and operation of LLMs necessitate significant computational resources, including powerful servers and large data centers, thus creating Energy Consumption and Sustainability Issues. As the demand for AI continues to grow, it is essential to critically examine the sustainability implications of LLMs and explore avenues for mitigating their environmental footprint~\cite{khowaja2023chatgpt}. As their prevalence and capabilities increase, we anticipate a rise in energy consumption, thereby intensifying concerns about sustainability and environmental impact. LLMs not only consume energy but also cause carbon emissions through the electricity they use to power their computational demands on the traditional electric grid. The carbon footprint of LLMs varies depending on factors such as electricity source and data center efficiency. However, studies have shown that the carbon emissions associated with training and operating LLMs can be significant, particularly when powered by fossil fuels. Given the urgency of addressing climate change, it is crucial to consider the environmental consequences of LLMs and explore ways to minimize their carbon footprint, underscoring a comprehensive approach to sustainability in AI development.

%\vspace{-2pt}

Bias in HAR can happen if a system treats some groups of people unfairly and on purpose, which can lead to wrong or biased results that affect decision making process. For example, HAR applications in healthcare can be affected by bias, which needs attention from individual, institutional, sectoral, and societal levels of efforts to mitigate~\cite{MORLEY2020113172}. The ethical AI aims to develop, deploy, and use HAR technologies in a manner that upholds ethical standards, respects human values, and mitigates potential harms by incorporating fairness, transparency, accountability, and privacy throughout the HAR lifecycle. It encompasses considerations such as avoiding bias in data collection, ensuring transparency in the decision-making processes of HAR algorithms, protecting user privacy and data rights, and promoting inclusivity and diversity in HAR development and deployment. Responsible AI goes beyond ethical considerations, encompassing the broader societal impacts and consequences of HAR technologies to contribute positively to society, minimize risks, and eliminate negative externalities. Together, ethical and responsible AI provide guidance for HAR technologies that prioritize ethical considerations, respect human rights, and society's well-being.

\section{Conclusion}\label{sec:conclusion}

The proposed framework for multidimensional analysis with LLM can prove effective and robust by examining data across various dimensions. Furthermore, steam processing of ETL tools can provide near-real-time data for stakeholders to gain a comprehensive understanding of HAR's capabilities, especially for older people and emergency situations. Additionally, we can use stored data to retrain GPT models, thereby enhancing the performance of LLM over time. However, the successful integration of LLM into these operations requires careful consideration of ethical, technical, and operational challenges, including data privacy, model interpretability, and human-machine interaction. Furthermore, data collection and processing have numerous challenges, like security, privacy, sustainability, and different cyber attacks that may affect decision-making. By removing these obstacles in multidimensional analysis, stakeholders can use the complex landscape with greater confidence and trust and transparency in AI systems across HAR applications.

\bibliographystyle{IEEEtran}
\bibliography{references}
\end{document}